%% file: paper.tex
\documentclass[runningheads,dvipsnames]{llncs}
\usepackage[T1]{fontenc}
\usepackage[utf8]{inputenc}
\usepackage{amsmath}
\usepackage{csquotes}
\usepackage{pgfplots}
\usepackage{cleveref}
\usepackage{multirow}
\usepackage{graphicx}
\usepackage{booktabs}
\usepackage{threeparttable}
\usepackage{tabularx}
\usepackage[hyphens]{url}
\usepackage{enumitem}
\newcommand*\circled[1]{\tikz[baseline=(char.base)]{
            \node[shape=circle,draw,inner sep=1pt] (char) {#1};}}

\def\makeheadbox{{%
\hbox to0pt{\vbox{\baselineskip=10dd\hrule\hbox
to\hsize{\vrule\kern3pt\vbox{\kern3pt
\hbox{\copyright Springer 2020.}
\hbox{The final authenticated version is available online at:}
\hbox{\url{https://doi.org/10.1007/978-3-030-48340-1_40}.}
\kern3pt}\hfil\kern3pt\vrule}\hrule}%
\hss}}}

\begin{document}

\makeheadbox

\title{Hugo: A Cluster Scheduler that Efficiently Learns to Select Complementary Data-Parallel Jobs}
% named in honor of the great scientific matchmatching methods proposed by Hugo Gernsbacher in the 1920s hahah: https://www.smithsonianmag.com/history/ mechanical-matchmaking-the-science-of-love-in-the-1920s-103877403/

\titlerunning{Hugo}
\author{Lauritz Thamsen\inst{1}, Ilya Verbitskiy\inst{1}, Sasho Nedelkoski\inst{1}, Vinh Thuy Tran\inst{1}, Vinícius Meyer\inst{2}, Miguel G. Xavier\inst{2}, Odej Kao\inst{1}, and César A. F. De Rose\inst{2}}
\authorrunning{Authors}
\institute{
TU Berlin, Germany, \{firstname.lastname\}@tu-berlin.de
\and
PUCRS, Brazil, \{firstname.lastname\}@pucrs.br
}

\pagestyle{plain}

\maketitle              % typeset the header of the contribution
\input{sections/0_abstract}
\input{sections/1_introduction}
\input{sections/2_related_work}
\input{sections/3_approach}
\input{sections/4_evaluation}
\input{sections/5_conclusion}
\input{sections/6_acknowledgments.tex}

\bibliographystyle{splncs04}
\bibliography{bib/bibliography}

\end{document}

%% file: sections/0_abstract.tex
\begin{abstract}

Distributed data processing systems like MapReduce, Spark, and Flink are popular tools for analysis of large datasets with cluster resources.
Yet, users often overprovision resources for their data processing jobs, while the resource usage of these jobs also typically fluctuates considerably.
Therefore, multiple jobs usually get scheduled onto the same shared resources to increase the resource utilization and throughput of clusters.
However, job runtimes and the utilization of shared resources can vary significantly depending on the specific combinations of co-located jobs.\\
This paper presents \emph{Hugo}, a cluster scheduler that continuously learns how efficiently jobs share resources, considering metrics for the resource utilization and interference among co-located jobs.
The scheduler combines offline grouping of jobs with online reinforcement learning to provide a scheduling mechanism that efficiently generalizes from specific monitored job combinations yet also adapts to changes in workloads.
Our evaluation of a prototype shows that the approach can reduce the runtimes of exemplary Spark jobs on a YARN cluster by up to 12.5\%, while resource utilization is increased and waiting times can be bounded.

\end{abstract}

%% file: sections/1_introduction.tex
\section{Introduction} \label{sec:INTRODUCTION}

Distributed data-parallel processing systems such as MapReduce~\cite{Dean_MapReduce_2004}, Spark~\cite{Zaharia_Spark_2010}, Flink~\cite{Carbone_Flink_2015}, and Dataflow/Beam~\cite{Akidau_Dataflow_2015} enable users to take advantage of clusters of bare-metal or virtual machines for analysis of large datasets.
These systems have become popular tools for workloads that range from data aggregation and search to relational queries, graph processing, and machine learning~\cite{Olston_Pig_2008,Gonzalez_GraphX_2014,Ghoting_SystemML_2011,Meng_MLlib_2016}.
Jobs from these diverse domains stress different resources, while the resource demands typically also fluctuate significantly over the runtime of jobs~\cite{Ousterhout_MSPDAF_2015,Renner_NeAwa_2015,Niu_Gemini_2015}.
Therefore, multiple jobs usually share cluster resources without isolation, so they can benefit from statistical multiplexing~\cite{Zaharia_FairScheduler_2010,Reiss_GTA_2012,Thamsen_LearningColocations_2018}.
This is implemented by using resource management systems like YARN~\cite{Vavilapalli_Yarn_2013} and Mesos~\cite{Hindman_Mesos_2011}.
These systems allow users to reserve fractions of cluster nodes via the notion of containers, in which users then run one or multiple jobs using the frameworks of their choice.
By default the resource management systems use simple scheduling methods such as round-robin, FIFO, greedy approaches, and other reservation-based methods such as dominant-resource fairness~\cite{Zaharia_Spark_2010,Hindman_Mesos_2011,Vavilapalli_Yarn_2013,Bao_ParameterScheduler_2018}, while low resource utilization remains a major problem in industry~\cite{Reiss_GTA_2012,Delimitrou_Quasar_2014,Rasley_EQM_2016}.
Yet, since jobs differ considerably in which resources they stress and how much utilization fluctuates, schedulers should actively co-locate jobs that share resources efficiently.
The benefits of such approaches have been demonstrated before, including by the authors~\cite{Thamsen_SRDD_2017,Ludwig_CIAPA_2018}, with multiple schedulers that explicitly take combined resource utilization and interference among co-located workloads into account~\cite{Delimitrou_Paragon_2013,Delimitrou_Quasar_2014,Niu_Gemini_2015} or learn the impact of this indirectly~\cite{Mao_DeepRM_2016,Mao_LearningScheduling_2018}, taking advantage of the recurrence of a majority of jobs~\cite{Jyothi_Morpheus_2016}.
However, previous efforts fall short in at least one of the following dimensions:
\begin{itemize}
    \item \emph{Learning Efficiency:} Multiple systems require extensive training data as they learn the sharing efficiency on the level of individual jobs or using completely generic learning methods~\cite{Niu_Gemini_2015,Thamsen_SRDD_2017,Mao_DeepRM_2016,Mao_LearningScheduling_2018}.
    \item \emph{Continuous Learning:} Some systems do not update their models continuously and therefore do not adapt to changes in workloads~\cite{Delimitrou_Paragon_2013,Ludwig_CIAPA_2018}.
    \item \emph{Solution Practicality:} Some systems do not incorporate objectives besides throughput~\cite{Mao_DeepRM_2016,Thamsen_SRDD_2017,Mao_LearningScheduling_2018}, while others assume control over more than just job order~\cite{Delimitrou_Paragon_2013,Delimitrou_Quasar_2014,Mao_LearningScheduling_2018} or require instrumentation not generally supported~\cite{Ludwig_CIAPA_2018}.
\end{itemize}

Addressing these limitations, we present Hugo, a cluster scheduler that efficiently learns from collected resource usage metrics to co-locate those jobs that have complementary resource demands and therefore share resources efficiently, building on our previous work~\cite{Thamsen_SRDD_2017,Ludwig_CIAPA_2018}.
Hugo first clusters jobs by their resource utilization, yielding multiple groups of jobs that contain jobs with similar resource demands.
Subsequently, our scheduler uses reinforcement learning to continuously evolve its knowledge on which groups of jobs are sharing the resources of a particular cluster environment efficiently.
That is, the scheduler learns for each workload and cluster from the experiences of scheduling particular job combinations onto the same cluster nodes, assessing which groups of jobs produce a high resource utilization yet low interference when co-located.
This combination of generalization across a fixed number of groups of jobs with reinforcement learning of co-location benefits provides learning efficiency, a reduced scheduling complexity, and adaptation to changes in workloads.
Furthermore, we show how additional scheduling requirements are integrated into Hugo with the example of balanced waiting times.

\paragraph{Contributions}
The contributions of this paper are:
\begin{itemize}
    \item We propose the scheduler \emph{Hugo}, which efficiently learns how different groups of cluster jobs of a data processing workload utilize shared resources.
    \item We implemented a prototype of our approach as a job submission tool for Spark jobs in YARN clusters.
    \item We evaluated our prototype on a cluster with 34 nodes, using different workloads and in comparison to YARN's default scheduling.
\end{itemize}

\paragraph{Outline}
The remainder of the paper is structured as follows.
Section~\ref{sec:RELATED_WORK} discusses the related work.
Section~\ref{sec:APPROACH} explains our scheduling approach.
Section~\ref{sec:EVALUATION} presents our evaluation of our approach.
Section~\ref{sec:CONCLUSION} concludes this paper.

%% file: sections/2_related_work.tex
\section{Related Work} \label{sec:RELATED_WORK}

In this section we describe related work on scheduling distributed data-parallel workloads based on resource utilization and interference.

Paragon~\cite{Delimitrou_Paragon_2013} profiles incoming jobs and matches them with jobs that are similar with regard to the impact of different hardware and interference with co-located workloads.
Paragon then assigns jobs to available resources using its job classes and collaborative filtering, aiming to minimize interference and maximize resource utilization.
In comparison, Hugo targets distributed data-parallel jobs and employs more resource utilization metrics for its co-location goodness.

Quasar~\cite{Delimitrou_Quasar_2014} uses classification to assess the impact of resources and interference with co-located workloads when scheduling jobs.
It takes performance requirements of users into account, monitors job performance at runtime, and adjusts models and allocations dynamically.
Quasar does assume full control over both resource allocation and assignment, while Hugo's scope is only scheduling of distributed data-parallel jobs.

Gemini~\cite{Niu_Gemini_2015} uses a model that captures the tradeoff between performance improvement and fairness loss for jobs scheduled in shared clusters.
The model quantifies the complementarity in the resource demands of jobs and is trained on historic workload data.
Gemini then decides automatically whether the fairness loss of a computed schedule is valid under a user's setting of required fairness and in relation to Dominant Resource Fairness~\cite{Ghodsi_DRF_2011}.
In comparison, Hugo uses a reinforcement learning algorithm and groups of jobs.

DeepRM~\cite{Mao_DeepRM_2016} is a scheduler that relies on deep reinforcement learning.
The scheduler models the state of a cluster system, taking into account the already allocated resources along with resource profiles for the queued jobs.
It then uses a neural network to obtain a probability distribution over all possible scheduling actions, using the rewards obtained after every action to update the parameters of the neural network.
In comparison, DeepRM is a generic reinforcement learning framework for job scheduling, whereas Hugo particularly targets job co-location effects.
Consequently, DeepRM might require more training effort.

Similar to DeepRM, Decima~\cite{Mao_LearningScheduling_2018} also uses reinforcement learning along with a neural network for job scheduling.
The system focuses on dataflow jobs that are described as directed acyclic graphs.
Decima does not only perform scheduling, but also learns job parameters such as task parallelism.
In comparison, Hugo does not make assumptions about a job's structure and configuration options.

We used reinforcement learning to co-locate cluster jobs based on the resource usage and interference before, including the same measure of co-location goodness~\cite{Thamsen_SRDD_2017}.
However, the previous approach maintains preferences of individual jobs, while Hugo learns and schedules on the level of job groups for efficiency and scalability.
Other related previous results include CIAPA~\cite{Ludwig_CIAPA_2018} and IntP~\cite{Xavier_Thesis_2018}.
CIAPA uses an interference- and affinity-aware performance model based on detailed system-level metrics to improve the placement of jobs.
IntP is a profiling tool that extracts fine-grained resource metrics from hardware counters and system structures.
Both these approaches use classification to generalize from individual jobs similar to Hugo's job groups.

%% file: sections/3_approach.tex
\section{Approach} \label{sec:APPROACH}

Hugo is an adaptive cluster job scheduler that utilizes resource usage profiles of jobs to select and co-locate combinations of jobs that efficiently share the available resources. It combines offline clustering and online reinforcement learning for efficient learning, scalability, and adaptation to changes in workloads.

\begin{figure}[htbp]
\centerline{\includegraphics[scale=0.55]{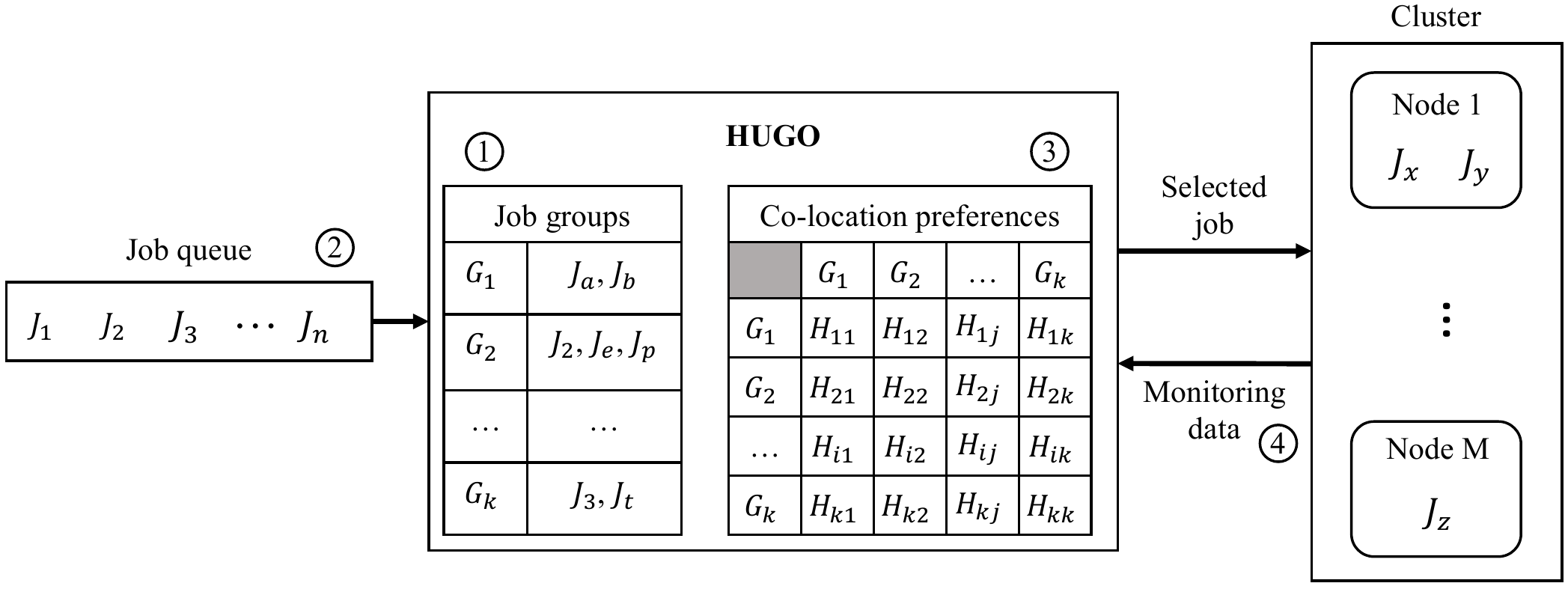}}
\caption{Scheduling jobs onto shared cluster resources based on continuously learned co-location preferences among groups of jobs.}
\label{generalapproach}
\end{figure}

Figure \ref{generalapproach} shows an overview of Hugo's approach.
It consists of the following main steps as annotated by the numbers in the figure:
\begin{enumerate}[label=\protect\circled{\arabic*}]
    \item Resource usage metrics of historic job runs are used to group jobs. For instance, two groups could distinguish CPU- and I/O-intensive jobs. The groups can be computed offline and can be updated periodically. Clustering methods can be used to establish a number of distinct groups automatically. Grouping provides an abstraction for the scheduling algorithm which then operates on job group level.
    \item Each incoming job is assigned to the job group it matches best. For this, either a profiling run on samples of the input or available historic monitoring data is used to match the job to one group.
    \item Using the groups, Hugo forms and continuously updates a preference matrix that quantifies the co-location goodness for pairs of job groups. The measure of co-location goodness is based on metrics that capture the combined resource utilization and the interference of co-located jobs.
    \item Based on the preference matrix, the scheduler selects those jobs from the queue that have a high co-location goodness with the currently running jobs in the cluster. The selected jobs are then scheduled onto the available resources.
    After a job is executed, the preference matrix is updated for the combination of jobs that were co-located via reinforcement learning.
\end{enumerate}

Utilizing reinforcement learning allows to learn the efficiency of different schedules during and through actual scheduling decisions. That is, scheduling decisions can be made right away on the basis of preferences, yet the scheduler learns continuously and, therefore, adapts to changes.

\subsection{Grouping Jobs based on Resource Metrics}

Grouping of the jobs is the key idea to ensure scalability and improve the learning efficiency of the scheduler, since it reduces the size of the preference matrix to $k$ groups.
Considering that we already have job resource usage statistics for all previously executed jobs we group them into $k$ groups.
Depending on the clustering method, and the job profiling information, the groups can have different meaning.
For example, there could be groups for jobs that predominantly stress the CPU, memory, disks, or network, while jobs of others groups could also exhibit mixed high usage of multiple resources such as both, CPU and memory.

Once we have the initial groups formed using the historic data for a specific cluster workload, the jobs in the queue need to be assigned to these groups.
Jobs in the queue can be recurring jobs or new jobs.
For recurring jobs we can use the previously recorded monitoring data to match the job to one of the available $k$ groups, for instance by averaging the utilization metrics of recent previous runs.
For new jobs, a profiling run is executed on a small sample of the entire input data to collect resource usage metrics and match the jobs to the available groups.
Using the resource usage profiles of the queued jobs each job is assigned to its representative group.

\subsection{Learning to Schedule Job Combinations}

We learn the co-location goodness on the level of job groups.
Therefore, our preference matrix contains a goodness measure for pairs of job groups.
The co-location goodness measure assesses how specific combinations of job groups utilize resources, using metrics that capture the resource utilization and interference among co-located jobs.
We use the same measure of co-location goodness and reinforcement learning algorithm we proposed previously for cluster scheduling~\cite{Thamsen_SRDD_2017}.

Typically, we have multiple jobs queued and with these jobs also multiple job groups.
Simultaneously, there are jobs already running on the shared cluster. We use reinforcement learning to select the jobs for scheduling and updating the preferences in the matrix based on the currently running as well as the queued jobs.
In the following we explain the job selection and the updates, which depend on each other.
We denote the preference matrix as $H$.
Its elements $H_{eg}$ contain the preference of job group $e$ when co-locating jobs of it with jobs of group $g$.
The probability of picking job group $g$ to run concurrently with an already running job of group $e$ is denoted as $\pi_{e}(g) = \frac{\exp({H_{eg}})}{\sum_{b\in S}\exp({H_{bg}})}$.
The probability of choosing a job group $g$ to select and schedule next on the cluster then is
\begin{equation*}
\Pi(g) = \frac{\sum_{e \in C}\Pi_{e}(g)}{\sum_{i \in Q}\sum_{e \in C}\Pi_{e}(i)}
\end{equation*}
where $\Pi_{e}(g) = \frac{\pi_{e}(g)}{\sum_{i \in Q}\pi_{e}(i)}$, $C$ is the set of job groups with jobs currently running on the cluster, $Q$ is the set of job groups with jobs currently in the queue and $S$ is the set of all groups.

To represent the relative goodness among the set $Q$, the probability for each job group $\Pi_{e}(g)$ is normalized. Job group $C_{\mathrm{next}}$ is chosen to be scheduled next by sampling from a distribution where the probabilities of each group are proportional to their co-location goodness with the existing job groups on the cluster.

After $C_{\mathrm{next}}$ is chosen, the scheduler scans the waiting queue and picks the jobs assigned to $C_{\mathrm{next}}$. If there is only one job match, it is scheduled next. Otherwise, if there are multiple options, a randomize function is used to choose one of the jobs in the group to schedule next.  Finally, for every node $n$ in the cluster the preferences between job groups of the co-location job pairs are subsequently updated as follows:
\begin{equation*}
H_{ij} := \alpha(R_{n} - \overline{R}^{i})\big(1 - \pi_{i}(j)\big) - \sum_{a\in \Omega_{n}\setminus\{i,j\}} \alpha(R_{n} - \overline{R}^{i})\pi_{i}(a) \qquad\forall i, j \in \Omega_{n}
\end{equation*}
where $\alpha$ is the learning rate, $\Omega_{n}$ is the set of job groups containing jobs placed on the node $n$, $R_{n}$ represents the co-location goodness for node n, and $\overline{R}^{i}$ is the mean goodness across all nodes containing jobs in the job group $i$.

\subsection{Integrating Additional Scheduling Constraints}

In practice, there are other requirements for scheduling beyond resource utilization and throughput.
Examples of these are fairness among users and priorities with particularly critical jobs. To highlight the practical applicability of Hugo we integrated a mechanism that balances waiting times and prevents job starvation.
As explained, when a job group is chosen as the next one to be scheduled, Hugo randomly chooses among the currently waiting jobs of that group with equal probabilities. To balance waiting times the choosing probabilities can be modified by the waiting times. The probabilities of choosing a job $a$ within the chosen job group $G$ is subsequently calculated by
\begin{equation*}
\pi_{a} = \frac{w_{a}}{\sum_{i\in G} w_i} 
\end{equation*}
where $w_{i}$ is the waiting time of job $i$.

The above usage of job waiting time only takes effect if the job group is eventually selected to schedule next. However, if the co-location goodness preference of the job group itself is low compared to most of the other job groups, the jobs in that group are still at risk of not getting selected.
We therefore further define the global parameter \emph{waiting limit} for the scheduler. When a job's waiting time reaches this limit, it is scheduled regardless of the co-location preferences. In case of multiple jobs with waiting times above the limit, one of them is chosen with probabilities according to the waiting time using the formula as above. We denote this version of the algorithm as Hugo*.

%% file: sections/4_evaluation.tex
\section{Evaluation} \label{sec:EVALUATION}

We tested our approach with four experiments using a prototype implementation, a commodity cluster, and various exemplary jobs.
In the following sections we describe the prototype implementation, the cluster setup, the test workload, and the four experiments along with the respective results.

\subsection{Prototype Implementation}
We implemented Hugo in Python as a job submission tool for YARN.
We use the combination of Telegraf and InfluxDB to monitor and persist the CPU, memory, disk, and network usage as well I/O wait data of each node.
Spark is chosen as the data processing engine. 
All the benchmarking jobs are implemented with Spark's APIs.

\subsection{Cluster Setup}
The experiments were executed on a cluster consisting of 34 nodes.
Each node is equipped with an Intel Xeon E3-1230 V2 @ 3.30GHz (4 physical cores, 8 virtual cores) and 16 GB of RAM, connected through a 1 Gbit/s Ethernet connection and running a Linux-based OS (kernel version 4.15.0).

Among the 34 nodes, one server was used to run YARN's resource manager.
YARN was configured such that each container occupied one logical CPU and 1800 MB of RAM, resulting in maximum of 8 containers per node.
From the 33 nodes managed by YARN, one node was used to run Spark's driver programs, while the remaining 32 nodes were used as worker nodes.

\subsection{Test Workload}

For simulating a mixed data processing workload, nine Spark analytic jobs are used throughout the experiment. The jobs and their datasets used for benchmarking are specified in \Cref{tbl:3_workload}.
For further reference each job is annotated by its own letter, while a number the annotated number denotes the job's group.
We grouped the jobs into six distinct groups by their utilization of CPU, disk, and memory, including groups of mixed utilization and overall low resource utilization.
The sizes of the input data are chosen so that the runtime of all jobs is similar and lasts approximately ten minutes.
The jobs were chosen such that they cover different application domains like machine learning (A, D, E, F),  graph processing (B, C),  relational queries (G),  and text processing (H, I).

\begin{table}[htb]
\centering
\caption{Dataflow jobs used in the experiment}\label{tbl:3_workload}
\begin{threeparttable}
\begin{tabularx}{\textwidth}{@{}p{36mm} l X@{}}
\toprule
Job (job, group) & Data source & Data parameters \\
\midrule
K-Means (A, 1)              & KMeansDataGenerator\tnote{\emph{a}}             & 100,000,000 points,\newline 80 clusters            \\
PageRank (B, 3)             & Graph Challenge data sets\tnote{\emph{b}}       & 46,656,000 edges,\newline 2,174,640 vertices         \\
Connected Components (C, 6) & Graph Challenge data sets\tnote{\emph{b}}       & 38,880,000 edges,\newline 1,812,200 vertices         \\
Linear Regression (D, 1)    & LinearDataGenerator\tnote{\emph{a}}             & 90,000,000 samples,\newline 20 features per sample \\
Logistic Regression (E, 2)  & LogisticRegressionDataGenerator\tnote{\emph{a}} & 11,000,000 samples,\newline 10 features per sample \\
SVM (F, 2)                  & SVMDataGenerator\tnote{\emph{a}}                & 70,000,000 samples,\newline 10 features per sample \\
TPC-H (G, 4)                & DBGEN\tnote{\emph{c}}                           & 100 GB generated DB                      \\
Sort (H, 5)                 & DBGEN\tnote{\emph{c}}                           & 143,999,787 records                        \\
Word Count (I, 1)           & Wikipedia backup data\tnote{\emph{d}}           & 53 GB text document                      \\
\bottomrule
\end{tabularx}
\begin{tablenotes}
    \item[\emph{a}] from the \texttt{org.apache.spark.mllib.util} package
    \item[\emph{b}] Graph Challenge datasets provided by Amazon, \url{https://graphchallenge.mit.edu/data-sets}
    \item[\emph{c}] included in TPC-H tool package, \url{http://www.tpc.org/tpch/}
    \item[\emph{d}] Wikipedia database dump, \url{https://dumps.wikimedia.org}
\end{tablenotes}
\end{threeparttable}
\end{table}

\subsection{Experiments}
We conducted four experiments.
Each experiment shows how our scheduler performs in a different scenario in comparison to the baseline round-robin scheduler in terms of makespan and resource utilization.
In the following we describe and motivate each of the experiments and present the results.

\subsubsection{Learning Phase}
The aim of this experiment is to gain insights into how our Hugo scheduler compares to the baseline round-robin scheduler when there is no preference data available when the scheduler starts.
That is, we start with an empty preference matrix.
The algorithm then populates and updates the preference matrix, continuously evaluating the resource usage of pairs of jobs.
To speed up the learning process, the job queue contains a job from each job group.
The jobs are placed in a repeating pattern as follows: C B G A F H $\times$ 10.

\paragraph{Results}
Using the Hugo scheduler all queued jobs took 169m 13s to finish as opposed to the round-robin scheduler with 180m 40s, an improvement by 6.3\%.
This result indicates that using the Hugo scheduler is beneficial in comparison to the round-robin scheduler, even without any prior preference data and therefore while training, when the workload contains periodically recurring jobs.

\subsubsection{Prior Preference Data}
In the follow-up experiment, the preference matrix output from the first experiment is used as the input for the Hugo scheduler.
However, in this experiment we exchange some of the jobs in the queue with jobs that did not appear in the queue of the previous experiment.
This way, we want to evaluate how well the grouping of our scheduler generalizes to unseen jobs.
The jobs are placed in a repeating pattern as follows: D E B C H G I $\times$ 5.

\paragraph{Results}
The Hugo scheduler again yields a faster running time: 114m 49s compared to the running time of the round-robin scheduler of 128m 17s. It thus produces a schedule that needs 10.5\% less time to finish all jobs.
Considering the job queue in this experiment has the same diversity of job groups as in the first experiment, the result suggests that the improvement is due to the prior knowledge of preferences between job groups.
Also, the result indicates that it is possible to have beneficial co-location of new jobs on the basis of the calculated co-location goodness of previously executed similar jobs.

\subsubsection{Randomized Queues}
In this experiment, we included all nine dataflow jobs.
The output preference matrix from the previous experiment is used as input preference data for our Hugo scheduler in this experiment. 
With this experiment we want to gain insights into how our scheduler behaves with a more realistic queue as opposed to the manually created queues of the previous experiments.
For this, we generated the following two randomized job queues: 
 C B B E A E E B I H H C B I H C E G F F A F C I G D A G I C G A F F D E G D A I D B H D H (Queue 1) and  E I A B C H G C A H E G C B F F G D B A C G D D H F I G C D B A F I F E I E E A H H B D I (Queue 2).

\paragraph{Results}
The results of this experiment are summarized in \Cref{tbl:3_randomized}.
Our Hugo scheduler does not only improve the utilization for each of the resources but also results in an improvement of up to 12.42\% in total processing time over the baseline round-robin scheduler.
This experiment, again, indicates that our scheduling approach is capable of finding advantageous co-locations that yield shorter execution times.

\begin{table}[h!]
\centering
\caption{Queue processing time for randomized queues}\label{tbl:3_randomized}
\begin{tabular}[t]{@{}llrrrrrr@{}}
\toprule
Queue               & Scheduler & CPU [\%] & Mem [\%] & Disk [\%] & Net [\%] & Duration & $\delta$ [\%]\\
\midrule
\multirow{2}{*}{1}  & RR &10.44&38.52&16.86&22.65 & 163m 15s & -- \\
                    & Hugo &12.45&40.74&17.36&28.85 & 142m 58s & -12.42 \\
\midrule
\multirow{2}{*}{2}  & RR &10.90&39.64&17.14&23.18 & 158m 48s & -- \\
                    & Hugo &12.84&41.46&18.54&25.33 & 141m 06s & -11.14 \\
\bottomrule
\end{tabular}
\end{table}

\subsubsection{Online Job Arrival}
The primary goal of the previous experiment was to assess whether our Hugo scheduler succeeds in placing those jobs onto shared nodes that run well together.
With the main focus being the co-location quality, however, the experiments disregarded the waiting times of jobs in the queue.
In this experiment, we evaluate how effective our extended scheduler,  Hugo*, deals with job queues where jobs have different waiting times.

The preference input data for this experiment is the output of the second experiment.
The job queue used for this experiment is the same randomized Job Queue 1 from the previous experiment. 
However, in contrast to the previous experiments, the whole queue is not known to the scheduler right away.
Instead, jobs join the queue after every scheduling round.
We look at two arrival patterns.
With constant arrival rate (CAR), a single job is added to the queue after every scheduling round.
With arbitrary arrival rate (AAR), 1 to 3 new jobs are added to the queue after every scheduling round.
The exact amount of jobs added to the queue follows a probability distribution where the probability of adding 1, 2, or 3 jobs equals 60\%, 20\%, and 20\%, respectively.

\paragraph{Results}

\begin{table}[h!]
    \caption{Queue processing time with arbitrary arrival rate}\label{tbl:3_waiting_times}
    \centering
    \begin{tabular}[t]{@{}lrr@{}}
    \toprule
    Scheduler & Duration  & $\delta$ [\%]\\
    \midrule
    RR        & 163m 15s & --\\
    Hugo* (CAR) & 145m 44s & -10.73\\
    Hugo* (AAR) & 154m 29s & -5.37\\
    \bottomrule
    \end{tabular}
\end{table}

\Cref{tbl:3_waiting_times} summarizes the outcome of this experiment.
Our Hugo scheduler is faster than the baseline for every constellation.
However, we also see a significant drop in performance with AAR.
A trade-off has to be made between job starvation and efficient job order.
This demonstrates the pitfall of when the scheduler is not able to submit jobs onto the cluster as fast as they arrive.

\begin{figure}[h!]
    \begin{tikzpicture}
        \begin{axis}[
            width=\textwidth,height=110pt,
            ymajorgrids,
            xtick=data,
            xmin=-1,xmax=37,
            ymin=0,ymax=14,
            ylabel={Number of jobs},
            xlabel={Waiting time (in scheduling rounds)},
            xtick={0,2,...,36},
            ytick={0,2,...,14},
            ybar=0pt,
            bar width=2.5pt,
            ylabel near ticks,
            legend style={font=\scriptsize},
            label style={font=\scriptsize},
            tick label style={font=\tiny},
            tick pos=left
          ]
            \addplot[fill=black] coordinates {
                ( 0, 9)
                ( 1, 3)
                ( 2, 2)
                ( 3, 0)
                ( 4, 2)
                ( 5, 2)
                ( 6, 1)
                ( 7, 4)
                ( 8, 1)
                ( 9, 1)
                (10, 3)
                (11, 0)
                (12, 0)
                (13, 3)
                (14, 1)
                (15, 1)
                (16, 0)
                (17, 1)
                (18, 0)
                (19, 0)
                (20, 2)
                (21, 0)
                (22, 0)
                (23, 1)
                (24, 0)
                (25, 1)
                (26, 2)
                (27, 0)
                (28, 0)
                (29, 0)
                (30, 1)
                (31, 1)
                (32, 0)
                (33, 2)
                (34, 0)
                (35, 0)
                (36, 1)
            };
            \addplot[fill=white] coordinates {
                ( 0, 6)
                ( 1, 4)
                ( 2, 2)
                ( 3, 3)
                ( 4, 1)
                ( 5, 0)
                ( 6, 2)
                ( 7, 0)
                ( 8, 2)
                ( 9, 0)
                (10, 2)
                (11, 0)
                (12, 1)
                (13, 2)
                (14, 0)
                (15, 1)
                (16, 1)
                (17, 2)
                (18, 0)
                (19, 1)
                (20,12)
                (21, 3)
                (22, 0)
                (23, 0)
                (24, 0)
                (25, 0)
                (26, 0)
                (27, 0)
                (28, 0)
                (29, 0)
                (30, 0)
                (31, 0)
                (32, 0)
                (33, 0)
                (34, 0)
                (35, 0)
                (36, 0)
            };
            \legend{Hugo, Hugo*}
        \end{axis}
    \end{tikzpicture}
    \caption{Comparison of job waiting times between the Hugo and the Hugo* scheduler.}\label{fig:3_waiting_times_aar}
    \end{figure}
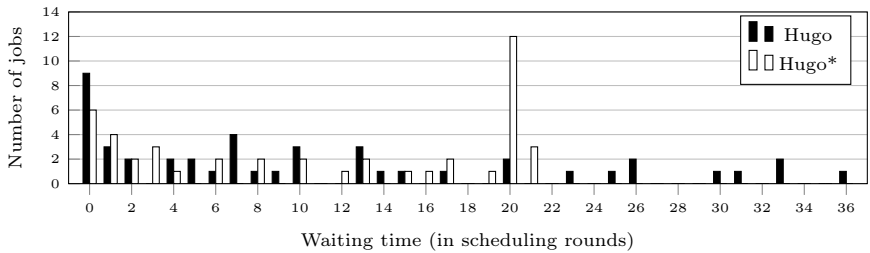

\Cref{fig:3_waiting_times_aar} shows how the waiting time is distributed with Hugo and Hugo*.
For Hugo there is a significantly higher number of jobs with or exceeding the waiting time limit of 20.
Hugo*, on the other hand, is able to successfully reduce the amount of jobs exceeding the waiting time limit.
However, since Hugo* gives jobs that are waiting longer than the global limit the highest preference, the schedules exhibit less optimal co-locations, reflected in longer total running time.

%% file: sections/5_conclusion.tex
\section{Conclusion} \label{sec:CONCLUSION}

This paper presented Hugo, a cluster scheduler for distributed data-parallel processing workloads that selects jobs based on the resource usage of co-located jobs.
Hugo uses a reinforcement learning algorithm to learn over time which combinations of jobs best utilize shared resources.
To efficiently generalize its knowledge and thus co-locate even new jobs effectively, the approach learns preferences not for single jobs but for groups of jobs that exhibit similar resource demands.
Hugo selects among the queued jobs using these learned preferences, choosing types of jobs that complement the jobs currently running on the shared infrastructure.
It thereby aims to schedule those jobs jointly onto shared nodes that yield the best overall resource utilization and runtimes.
We implemented a prototype of Hugo for Spark and YARN, showing that given mixed workloads with recurring jobs, our approach can reduce job runtimes, increase resource utilization, and still balance waiting times.

%% file: sections/6_acknowledgments.tex
\subsubsection*{Acknowledgments} \label{sec:ACKNOWLEDGMENTS}

This work has been supported through grants by the German Ministry for Education and Research (BMBF; funding mark 01IS14013A and 01IS18025A).